\documentclass[12pt,showkeys,amsmath,amssymb]{revtex4}
\usepackage{amsmath,amsfonts,amsthm,amscd,amssymb,latexsym}
\usepackage{bm}
\usepackage{dcolumn}
\usepackage{graphicx}
\usepackage{epstopdf}
\usepackage{color}
\usepackage{epsf}
\usepackage{epsfig}
\usepackage{graphicx, epic, eepic, color}
\usepackage[colorlinks,citecolor=blue,urlcolor=blue,linkcolor=blue]{hyperref}
\usepackage{float}
\usepackage[caption = false]{subfig}

\begin{document}

\title{Cosmic Jets in General Relativity}

\author{Bahram \surname{Mashhoon}$^{1,2,3}$}
\email{mashhoonb@missouri.edu}

\affiliation{$^1$Department of Physics, Sharif University of Technology, Tehran 11155-9161, Iran\\ 
$^2$School of Astronomy, Institute for Research in Fundamental Sciences (IPM), Tehran 19395-5531, Iran\\
$^3$Department of Physics and Astronomy, University of Missouri, Columbia,
Missouri 65211, USA\\
}

\begin{abstract}
In certain general relativistic time-dependent gravitational fields, free test particles can asymptotically line up relative to fiducial static observers and produce a cosmic jet whose speed approaches the speed of light. Two scenarios for the formation of these purely gravitational jets have thus far been theoretically demonstrated: the double-jet collapse scenario  and the wave scenario that involves both a cosmic jet in the general direction of wave propagation as well as a counterjet in the opposite direction. These scenarios  are briefly reviewed in this paper. Moreover,  to elucidate the process of jet formation in these scenarios, we study the propagation of scalar fields in related gravitational fields in connection with the correspondence between the scalar wave perturbations and the motion of free test particles and null rays.   
\end{abstract}


\maketitle

\section{introduction}

The main purpose of this paper is to describe theoretical jet-like structures that can occur in dynamic spacetimes of general relativity. Indeed, \emph{cosmic jets} are theoretical constructs that are purely gravitational in nature and are thus quite distinct from 
\emph{astrophysical jets} that are typically collimated outflows of ionized matter along the rotation axis of a gravitationally collapsed configuration. Astrophysical jets occur in many astronomical systems; moreover, jet formation involves electromagnetic and gravitational processes that are rather complex. The bipolar outflows associated with quasars and active galactic nuclei usually involve a pair of jets that originate from a central ``engine" surrounded by an accretion disk. In addition, most of the x-ray binary systems in our galaxy also exhibit the double-jet structure (``microquasars"). In relativistic astrophysical jets, beams of magnetized plasma are accelerated close to the speed of light; see, for instance,~\cite{MDS, Romero:2016hjn, Romero:2021bet, Hada:2024icg} and the references cited therein. 

In general relativity (GR), there are at present two known schemes for the formation of cosmic jets, namely, the collapse scenario and the wave scenario; these are described in Sections II and IV, respectively.  Is there a connection between the collapse scenario and the formation of astrophysical jets? The answer to this intriguing question is not known at the present time. The rest of this section is devoted to explaining how the cosmic jet phenomenon could come about in time-dependent gravitational fields in GR. 

Let us consider a spacetime with metric 
\begin{equation}\label{I1}
ds^2= g_{\mu \nu}(x) dx^\mu\,dx^\nu
\end{equation}
expressed in an admissible system of coordinates $x^\mu = (ct, x^i)$. We assume that this gravitational field is a solution of Einstein's field equations 
\begin{equation}\label{I2}
R_{\mu \nu} - \frac{1}{2}\,g_{\mu \nu} R + \Lambda \,g_{\mu \nu} = \kappa\, T_{\mu \nu}\,,
\end{equation}
where $\kappa = 8 \pi G/c^4$, $\Lambda$ is the cosmological constant and $T_{\mu \nu}$ is the energy-momentum tensor of the source of the gravitational field. In this paper, greek indices run from 0 to 3, while latin indices run from 1 to 3; moreover,  the signature of the metric is assumed to be $+2$. Henceforth, we set $c = 1$, unless otherwise specified.

Free test particles and null rays follow spacetime geodesics in GR. Let $u^\mu = dx^\mu/d\lambda$ be the 4-velocity vector of a geodesic world line; here, $\lambda$ is the proper time $\tau$ along a timelike geodesic or the affine parameter along a null geodesic. Moreover,  $u^\mu\,u_\mu = - \varepsilon$, where $\varepsilon$ is unity (zero) for timelike (null) geodesics.  In a stationary gravitational field, a timelike Killing vector $\xi^\mu$ exists such that the projection of the 4-velocity vector $u^\mu$ of a free test particle or null ray upon $\xi^\mu$, namely, $u^\mu\,\xi_\mu$ is a constant of the motion. This result has the interpretation that the free test particle or null ray does not exchange energy with the gravitational field as a consequence of the invariance of the background spacetime under translation in time. The situation is quite different, however, in a dynamic time-dependent gravitational field in GR. Under certain circumstances, free test particles can gain energy from the background dynamic gravitational field and thereby generate cosmic jets.   

To describe the formation of cosmic jets in a coordinate invariant manner, we introduce the congruence of the fundamental comoving observers that are all spatially at rest in the time-dependent background gravitational field under consideration here. Let
\begin{equation}\label{I3}
U^\mu = (- g_{tt})^{-\tfrac{1}{2}}\, \delta^{\mu}_{0}
\end{equation}
be the 4-velocity vector of the preferred observers and $\chi^{\mu}{}_{\hat \alpha}$ be their natural adapted tetrad frames such that $\chi^{\mu}{}_{\hat 0} = U^\mu$. The 4-velocity vector of a free test particle or null ray as measured by a fiducial comoving observer is then 
\begin{equation}\label{I4}
u^{\hat \alpha} = u^\mu\,\chi_{\mu}{}^{\hat \alpha}\,.
\end{equation}

We now turn to the rudiments of cosmic jet formation in the collapse and wave scenarios.

\section{Collapse Scenario}

To describe the collapse scenario in a straightforward setting, we consider the standard Ricci-flat Kasner solution of GR given by~\cite{Kasner:1921zz, WK}
\begin{equation}\label{C1}
ds^2 = - dt^2+t^{2p_1} dx^2+t^{2 p_2} dy^2+t^{2 p_3} dz^2\,,
\end{equation}
where the coordinate system is admissible for $t \in (0, \infty)$, 
\begin{equation}\label{C2}
p_1+p_2+p_3 = p_1^2+p_2^2+p_3^2 = 1\,
\end{equation}
and  $(-g)^{\frac{1}{2}} = t$. For $(p_1, p_2, p_3) = (1, 0, 0)$, the Kasner spacetime reduces to the Minkowski spacetime; moreover, for $(p_1, p_2, p_3) = (\tfrac{2}{3}, \tfrac{2}{3}, -\tfrac{1}{3})$, we have a plane-symmetric Petrov type D spacetime. Except for these and similar cases, the Kasner models are algebraically general and belong to Petrov type I. The standard Kasner spacetime thus represents a spatially homogeneous (Bianchi type I) but anisotropic cosmological 
model~\cite{Stephani:2003tm, Griffiths:2009dfa}. 
Despite their simplicity, the Kasner solutions play a significant role in the theoretical description of the behavior of generic cosmological solutions of GR on approach to the singularity~\cite{WK}.  

Henceforth, we assume that  $p_1 < p_2 < p_3$; that is, we choose
\begin{equation}\label{C3}
-\frac{1}{3}\le p_1\le 0\,,\qquad 0\le p_2\le \frac{2}{3}\,,\qquad \frac{2}{3}\le p_3\le 1\,,
\end{equation}
so that in Eq.~\eqref{C1} the proper distance along the $x$ axis decreases to zero asymptotically (i.e., for $t\to\infty$), while the proper distances along the $y$ and $z$ axes tend asymptotically to infinity. 

We are interested in the behavior of future-directed timelike and null geodesics in the standard Kasner metric~\cite{Chicone:2010xr}. It follows from the existence of the spacelike Killing vector fields  $\partial_x$,  $\partial_y$    and $\partial_z$ that we have three constants of the motion $k_1$, $k_2$ and $k_3$ that result from the projection of $u^\mu = dx^\mu/d\lambda$ upon the three Killing vectors, respectively. That is, 
 \begin{equation}\label{C4}
t^{2 p_1}\,\frac{dx}{d\lambda} = k_1\,,\qquad   t^{2 p_2}\,\frac{dy}{d\lambda} = k_2\,,\qquad t^{2 p_3}\,\frac{dz}{d\lambda} = k_3\,.
\end{equation}
Next, from $u^\mu \,u_\mu = - \varepsilon$, we find for the future-directed geodesic world lines
\begin{equation}\label{C5}
\frac{dt}{d\lambda} = (\varepsilon + W)^{\frac{1}{2}}\,, \qquad  W := k_1^2\, t^{-2 p_1} + k_2^2 \,t^{-2 p_2} + k_3^2\, t^{-2 p_3}\,.
\end{equation}
The explicit solution for the 4-velocity vector  of future-directed timelike and null geodesics is thus given by
\begin{equation}\label{C6}
u^\mu = \left[(\varepsilon + W)^{\frac{1}{2}}, k_1\, t^{-2 p_1}, k_2 \,t^{-2 p_2},  k_3\, t^{-2 p_3}\right]\,.
\end{equation}

In connection with preferred observers in the Kasner spacetime, let us mention a general theorem in GR which states that observers that remain spatially at rest in a gravitational field with a metric of the form $-dt^2 + g_{ij}(x)\, dx^i \,dx^j $ follow spacetime geodesics~\cite{Chicone:2002kb}. In the present case, we have $(k_1, k_2, k_3) = 0$ and $U^\mu = \delta^\mu_0$ for the 4-velocity vector of the fiducial observers; moreover, their natural adapted orthonormal tetrads $\chi^{\mu}{}_{\hat \alpha}$ are given by
$\chi^{\mu}{}_{\hat 0} = U^\mu = \delta^\mu_0$ and
\begin{equation}\label{C7}
\chi^{\mu}{}_{\hat 1} = (0,t^{-p_1},0,0)\,,\qquad \chi^{\mu}{}_{\hat 2} = (0, 0, t^{-p_2},0)\,,\qquad \chi^{\mu}{}_{\hat 3} = (0,0,0,t^{-p_3})\,.
\end{equation}
This tetrad frame field is parallel propagated along the geodesic world lines of the fiducial observers in Kasner spacetime. 

We are interested in the motion of free test particles and null rays as observed by the preferred observers in  the Kasner spacetime. Using Eq.~\eqref{I4}, we find 
\begin{equation}\label{C8}
u^{\hat \alpha} = \left[(\varepsilon + W)^{\frac{1}{2}}, k_1\, t^{- p_1}, k_2 \,t^{- p_2},  k_3\, t^{- p_3}\right]\,.
\end{equation}

\subsection{Free Test Particles}

For a free test particle, $\varepsilon = 1$, $\lambda = \tau$ and $dt/d\tau = \gamma$, where $\gamma$ is the measured Lorentz factor; that is, 
\begin{equation}\label{C9}
u^{\hat \alpha} = \gamma ( 1, \bm{\hat v})\,,
\end{equation}
where
\begin{equation}\label{C10}
\hat{v}_x = \frac{ k_1\, t^{- p_1}}{(1 + W)^{\frac{1}{2}}}\,, \qquad \hat{v}_y= \frac{k_2 \,t^{- p_2}}{(1 + W)^{\frac{1}{2}}}\,, \qquad \hat{v}_z = \frac{k_3\, t^{- p_3}}{(1 + W)^{\frac{1}{2}}}\,.
\end{equation}
As $t \to \infty$, we find 
\begin{equation}\label{C11}
\bm{\hat v} \to (\frac{k_1}{|k_1|}, 0, 0)\,,
\end{equation}
which indicates the formation of a double-jet configuration up ($k_1 > 0$) and down ($k_1 < 0$) along the collapsing direction, namely, the $x$ axis.  The speed of the jet asymptotically approaches the speed of light as measured by the background comoving observers. This comoving congruence would represent the ambient medium in an astrophysical context. 

For test particles with $k_1 = 0$, we find $\bm{\hat v} \to 0$ as $t \to \infty$. That is, in the expanding directions, the measured speeds tend to zero. 

In astrophysical jets, the bipolar outflows of ionized matter are emitted away from a central gravitationally collapsed configuration; therefore, one cannot help but wonder if there is a link between the collapse scenario and the formation of 
astrophysical jets~\cite{Chicone:2010xr}.  This remains an open problem. It may eventually turn out that the cosmic jets of the collapse scenario and the astrophysical jets are different aspects of the same basic natural phenomenon.

\subsection{Free Null Rays}

For null rays, $\varepsilon = 0$ and $\lambda$ is an affine parameter. We therefore have
\begin{equation}\label{C12}
\hat{v}_x = k_1\, t^{- p_1}\,W^{-\frac{1}{2}}\,, \qquad \hat{v}_y= k_2 \,t^{- p_2}\,W^{-\frac{1}{2}}\,, \qquad \hat{v}_z = k_3\, t^{- p_3}\,W^{-\frac{1}{2}}\,
\end{equation}
such that 
\begin{equation}\label{C13}
\hat{v}^2_x + \hat{v}^2_y + \hat{v}^2_z = 1\,.
\end{equation}
For $t \to \infty$, we find as before that the null geodesics line up along the collapse axis and 
\begin{equation}\label{C14}
\bm{\hat v}|_{\rm null} \to (\frac{k_1}{|k_1|}, 0, 0)\,.
\end{equation}
If $k_1 = 0$ but $k_2 \ne 0$, then 
\begin{equation}\label{C15}
\bm{\hat v}|_{\rm null} \to (0, \frac{k_2}{|k_2|}, 0)\,,
\end{equation}
etc. 

\subsection{$t \to 0$}

The conclusions we have reached for $t \to \infty$ are quite general and should apply equally well on approach to the singularity at $t = 0$. Indeed, as $t \to 0$, inspection of Eq.~\eqref{C1} reveals that expansion occurs in the Kasner metric along the $x$ axis and collapse occurs along the $y$ and $z$ axes; however, contraction is strongest along the $z$ direction. Therefore, Eq.~\eqref{C10} implies that for free test particles $\hat{v}_x \to 0$ along the expanding direction as $t \to 0$. On the other hand, if $k_3 \ne 0$, then 
\begin{equation}\label{C16}
\bm{\hat v}|_{t\to 0} \to (0, 0, \frac{k_3}{|k_3|})\,,
\end{equation}
while if $k_3 = 0$, then
\begin{equation}\label{C17}
\bm{\hat v}|_{t\to 0} \to (0, \frac{k_2}{|k_2|}, 0)\,.
\end{equation}
We thus have a double-jet configuration along a contracting direction such that the speed of the jet approaches the speed of light as $t \to 0$. 

For null rays, the results for $t \to 0$ turn out to be the same as for free test particles in the cases of Eqs.~\eqref{C16} and~\eqref{C17}, while in the case that $k_2 = k_3 = 0$, we have $\hat{v}^2_x= 1$.

It is interesting to mention here that the notion of cosmic jets in GR first came to light with the publication of two papers on gravitomagnetic jets~\cite{Chicone:2010aa, Chicone:2010hy}. Subsequent work involving various time-dependent solutions of GR have demonstrated the robust nature of the collapse scenario~\cite{Chicone:2010xr, Chicone:2011ie, Bini:2014esa, Bini:2017qnd, Bini:2022xzk}. Meanwhile, various interesting attempts have been made to make contact with the problem of astrophysical jets~\cite{Poirier:2015cyu, Poirier:2016rul, Tucker:2016wvt, Tucker:2018xle}. 

To gain further insight into the nature of the collapse scenario, we study the behavior of the scalar field equation in the Kasner spacetime. 


\section{Collapse Scenario: Scalar Field}

It is interesting to compare the behavior of geodesics in a spacetime with metric $g_{\mu \nu}$ with the propagation of matter waves represented by a scalar field $\Phi$ with inertial mass $m$. Let $\epsilon$ be the amplitude of the scalar wave; then,  the influence of this perturbation on the Kasner geometry under consideration here turns out to be of order $\epsilon^2$ and can be neglected. Thus we can consider the scalar field as a simple test field. The corresponding wave equation can be derived using the variational principle of stationary action, $\delta \int \mathcal{L}\,d^4x = 0$, where $\mathcal{L}$ is the Lagrangian density given by
\begin{equation}\label{S1}
\mathcal{L} = -\frac{1}{2} (-g)^{\frac{1}{2}} \left(g^{\mu \nu} \Phi_{,\mu} \Phi_{,\nu}  + \frac{1}{\ell^2}\,\Phi^2\right)\,.
\end{equation}
Here,  $\ell := \hbar /(mc)$ is the Compton wavelength of the particle.  The scalar wave equation is linear and can be written as
\begin{equation}\label{S2}
g^{\mu \nu} \Phi_{; \mu \nu} - \frac{1}{\ell^2}\Phi = 0\,.
\end{equation}
More explicitly, we have 
\begin{equation}\label{S3}
\frac{1}{(-g)^{\frac{1}{2}}}\,\frac{\partial}{\partial x^\mu} \left[(-g)^{\frac{1}{2}} \,g^{\mu \nu}\frac{\partial \Phi}{\partial x^\nu} \right]  - \frac{1}{\ell^2}\Phi = 0\,.
\end{equation}

In the WKB limit, this equation reduces to the timelike (null) geodesic equation for $m \ne 0$ ($m =0$). This circumstance is related to the connection established by the correspondence principle between wave mechanics and particle mechanics in the eikonal limit. 

For the Kasner universe, $(-g)^{\frac{1}{2}} = t$ and the scalar wave equation reduces to 
\begin{equation}\label{S4}
 -\frac{1}{t}\frac{\partial}{\partial t}\left(t \frac{\partial \Phi}{\partial t}\right) +t^{-2 p_1}\, \frac{\partial^2 \Phi}{\partial x^2} +t^{-2 p_2}\, \frac{\partial^2 \Phi}{\partial y^2}
 +t^{-2 p_3}\,\frac{\partial^2 \Phi}{\partial z^2} - \frac{1}{\ell^2} \Phi = 0\,.
\end{equation}
The Kasner metric is invariant under translations in $x$, $y$ and $z$ coordinates; hence, we can assume a solution of Eq.~\eqref{S4} of the form
\begin{equation}\label{S5}
 \Phi = e^{i  \kappa_1\, x + i  \kappa_2 \,y +i  \kappa_3\, z} \psi(t)\,,
\end{equation}
where $ \kappa_1$, $ \kappa_2$ and $ \kappa_3$ are constants. The scalar field $\Phi$ is a real function that represents a neutral particle of mass $m$ and satisfies a linear wave equation; nevertheless,  to simplify matters,  we can write $\Phi$ in  complex form~\eqref{S5} with the proviso that only its real part has physical significance.   It follows from Eqs.~\eqref{S4} and~\eqref{S5} that 
\begin{equation}\label{S6}
\frac{1}{t}\frac{d}{d t}\left(t\, \frac{d \psi}{d t}\right) + (\mathcal{W} + \ell^{-2})\psi = 0\,, \quad \mathcal{W} := \kappa_1^2\, t^{-2 p_1} +  \kappa_2^2 \,t^{-2 p_2} +  \kappa_3^2\, t^{-2 p_3}\,.
\end{equation}
This ordinary differential equation of second order has  singular points at $t = 0$ and $t = \infty$. The general solution for the scalar field is a linear superposition of solutions of the form given by Eqs.~\eqref{S5} and~\eqref{S6} for all possible values of the  constants $ \kappa_1$, $ \kappa_2$ and $ \kappa_3$.

It does not appear possible to find explicit solutions for Eq.~\eqref{S6}; therefore, we must content ourselves with some qualitative remarks.  The solution of Eq.~\eqref{S6} for the wave amplitude $\psi$ is dominated by the magnitude of $\mathcal{W}$; furthermore, it is remarkable that upon replacing $(\kappa_1,  \kappa_2,   \kappa_3)$ by $(k_1, k_2, k_3)$, $\mathcal{W}$ turns into $W$ that we introduced in Eq.~\eqref{C5} for future-directed Kasner geodesics and played a crucial role in the discussion of the collapse scenario. Indeed, for $\kappa_1 \ne 0$, the solution of the wave amplitude $\psi$ is dominated by the first term in $\mathcal{W}$ for $t \to \infty$, while the third term with $\kappa_3 \ne 0$ dominates for $t \to 0$. Let us note that the wave equation for $\psi$ depends, for instance, upon  $\kappa_1^2$; hence, the jets $\kappa_1 > 0$ and $ \kappa_1 < 0$ are treated in the same way in Eq.~\eqref{S6}. These results are qualitatively the same for free test particles and null rays, since our considerations are independent of the magnitude of $\ell$.


\section{Wave Scenario}

To illustrate the wave scenario in a simple setting, we resort to plane gravitational waves. Gravitational plane waves in GR form a subclass of plane-fronted gravitational waves with parallel rays ($pp$-waves) that were first discussed by Brinkmann~\cite{Brinkmann:1925fr} and have been subsequently studied by a number of authors~\cite{BaJe, Rosen, Bondi:1957dt, Bondi:1958aj, Ehlers:1962zz, Zhang:2017rno, Hogan:2022reu, Roche:2022bcz}. The $pp$-waves are Ricci-flat solutions of GR that admit a covariantly constant null vector field.
 A plane gravitational wave thus admits parallel null rays  $K^\mu$ such that $K^\mu\,K_\mu = 0$ and $K_{\mu;\nu} = 0$. Moreover, plane waves are of type N in the Petrov classification; that is, the four principal null directions of the Weyl tensor coincide in this case and indicate the direction of wave propagation $K^\mu$ that is normal to the uniform wave front. Furthermore, plane waves possess five Killing vector fields~\cite{Bondi:1958aj}. For a detailed discussion of solutions of GR that represent gravitational waves, see~\cite{Stephani:2003tm, Griffiths:2009dfa} and the references therein. 

For the sake of concreteness, we consider a plane gravitational wave spacetime of the form~\cite{Bini:2017qnd}
\begin{equation}\label{L1}
ds^2 = -dt^2 +dz^2 + \mathfrak{u}^{2s_1} dx^2 + \mathfrak{u}^{2s_2} dy^2\,, 
\end{equation}
where $s_1$ and $s_2$ are constants and $\mathfrak{u}$ is the retarded null coordinate
\begin{equation}\label{L2}
 \mathfrak{u} = t-z\,.
\end{equation}
The coordinate system is admissible once $\mathfrak{u} > 0$. Equation~\eqref{L1} represents an exact linearly polarized plane gravitational wave with plus (``$\oplus$") polarization~\cite{Bini:2017qnd}; moreover, this spacetime contains a null Killing vector field $K = \partial_t + \partial_z$ with $K_\mu = - \partial_\mu  \mathfrak{u}$.  The vector $K$ is the propagation vector of the wave in the $z$ direction at the speed of light and  constitutes a nonexpanding shearfree geodesic congruence.

Einstein's field equations in vacuum ($R_{\mu\nu}=0$) are satisfied provided
\begin{equation}\label{L3}
s_1^2 + s_2^2 = s_1 + s_2\,.
\end{equation}
We note that 
\begin{equation}\label{L4}
(s_1-\tfrac{1}{2})^2 + (s_2-\tfrac{1}{2})^2 = \tfrac{1}{2}\,.
\end{equation}
Therefore, 
\begin{equation}\label{L5}
s_1 =  \tfrac{1}{2} +  \tfrac{1}{\sqrt{2}}\,\cos \theta\,,\qquad s_2 =  \tfrac{1}{2} +  \tfrac{1}{\sqrt{2}}\,\sin \theta\,,
\end{equation}
where $\theta$, $0 \le \theta < 2 \pi$ is an arbitrary angle.

In this spacetime,  preferred observers at rest follow timelike geodesic world lines and their natural tetrad frames, namely,
\begin{equation}
\label{L6}
e_{\hat 0}=\partial_t\,,\quad e_{\hat 1} = \mathfrak{u}^{-s_1} \,\partial_x\,,\quad
e_{\hat 2} = \mathfrak{u}^{-s_2} \,\partial_y\,,\quad e_{\hat 3} = \partial_z\,,
\end{equation}
are parallel propagated along their geodesic world lines. 

In connection with the spacetime curvature in this plane wave spacetime, we assume that the preferred observers measure the spacetime curvature by projecting the Riemann tensor on their adapted tetrad frames, namely, 
\begin{equation}\label{L7}
R_{\hat \alpha \hat \beta \hat \gamma \hat \delta} = R_{\mu \nu \rho \sigma}\,e^{\mu}{}_{\hat \alpha}\,e^{\nu}{}_{\hat \beta}\,e^{\rho}{}_{\hat \gamma}\,e^{\sigma}{}_{\hat \delta}\,.
\end{equation}
We can express Eq.~\eqref{L7} as a $6\times 6$ matrix with indices that range over the set $\{01,02,03,23,31,12\}$. The result,  in any Ricci-flat spacetime, involves $3\times 3$ matrices in the form
\begin{equation}
\label{L8}
\mathcal{C} = \left[
\begin{array}{cc}
\mathcal{E} & \mathcal{B}\cr
\mathcal{B} & - \mathcal{E}\cr
\end{array}
\right]\,,
\end{equation}
where $\mathcal{E}$  and $\mathcal{B}$ are symmetric and traceless matrices and represent the \lq\lq electric" and \lq\lq magnetic" components of the Weyl conformal curvature tensor, respectively. For metric~\eqref{L1}, we find 
\begin{equation}
\label{L9}
{\mathcal E}= \mathbb{K}(\mathfrak{u})\left[
\begin{array}{ccc}
1&0 & 0\cr
0&-1 & 0\cr
0&0 & 0\cr
\end{array}
\right]\,,\qquad
{\mathcal B}= \mathbb{K}(\mathfrak{u})\left[
\begin{array}{ccc}
0&-1 & 0\cr
-1&0 & 0\cr
0&0 & 0\cr
\end{array}
\right]\,,
\end{equation}
where
\begin{equation}
\label{L10}
\mathbb{K}(\mathfrak{u}) = \frac{s_1(1-s_1)}{\mathfrak{u}^2} = \frac{s_2 (s_2-1)}{\mathfrak{u}^2} = - \frac{\cos 2\theta}{4 \,\mathfrak{u}^2}\,.
\end{equation}
The spacetime is flat if either $s_1$ or $s_2$ vanishes or is equal to unity; otherwise, there is a spacetime singularity at $\mathfrak{u} = 0$. 

The Killing vectors for this spacetime are $\partial_x$, $\partial_y$,  $\partial_t + \partial_z$ and 
\begin{equation}\label{L11}
x\,(\partial_t + \partial_z) + \frac{\mathfrak{u}^{1-2s_1}}{1-2s_1}\, \partial_x\,, \qquad y\,(\partial_t + \partial_z)  + \frac{\mathfrak{u}^{1-2s_2}}{1-2s_2}\, \partial_y\,
\end{equation}
for $s_1 \ne 1/2$ and $s_2 \ne 1/2$. Let us suppose, for instance, that $s_1 = 1/2$ and $s_2 = (1\pm \sqrt{2})/2$; then, the appropriate Killing vector in Eq.~\eqref{L11} must be replaced by 
\begin{equation}\label{L12}
x\,(\partial_t + \partial_z) + \ln{\mathfrak{u}}\, \partial_x\,.
\end{equation}

Regarding the wave scenario in the gravitational field represented by Eq.~\eqref{L1}, we consider future-directed timelike and null geodesics  with 4-velocity vectors  $v^\mu = dx^\mu/d\eta$, where $\eta$ is the proper time $\tau$ along a timelike geodesic world line or the affine parameter along a null geodesic. Here,  $v^\mu v_\mu = -\varepsilon$, where, as before,  $\varepsilon$ is unity (zero) for timelike (null) geodesics. The projection of $v^\mu$ on the Killing vector fields $\partial_t + \partial_z$, $\partial_x$ and $\partial_y$ are constants of the motion; hence, 
\begin{equation}\label{L13}
- \frac{dt}{d\eta} + \frac{dz}{d\eta} = -q_0\,, \qquad \frac{d\mathfrak{u}}{d\eta} = q_0\,,
\end{equation}
\begin{equation}\label{L14}
\mathfrak{u}^{2s_1}\,\frac{dx}{d\eta} = q_1\,, \qquad   \mathfrak{u}^{2s_2}\,\frac{dy}{d\eta} = q_2\,,
\end{equation}
where $q_0$, $q_1$ and $q_2$ are constants. From $v^\mu v_\mu = -\varepsilon$ and Eqs.~\eqref{L13} and~\eqref{L14}, we find 
\begin{equation}\label{L15}
q_0\,\frac{d(t+z)}{d\eta} =  \varepsilon + q^2_1\, \mathfrak{u}^{-2s_1}+ q^2_2\, \mathfrak{u}^{-2s_2}\,.
\end{equation}
If $q_0 = 0$, then the only consistent solution for $v^\mu$ is that it is proportional to $K^\mu$, the plane wave propagation vector. Therefore, we assume henceforward that $q_0 > 0$ for future-directed geodesics. It is then straightforward to show
\begin{equation}\label{L16}
\frac{dt}{d\eta} = \frac{1}{2}\left(\frac{\varepsilon}{q_0} + q_0\right) + \frac{1}{2q_0}\,(q^2_1\, \mathfrak{u}^{-2s_1}+ q^2_2\, \mathfrak{u}^{-2s_2})\,,
\end{equation}
\begin{equation}\label{L17}
 \frac{dz}{d\eta} = \frac{1}{2}\left(\frac{\varepsilon}{q_0} - q_0\right) + \frac{1}{2q_0}\,(q^2_1\, \mathfrak{u}^{-2s_1}+ q^2_2\, \mathfrak{u}^{-2s_2})\,.
\end{equation}

Let us note that with $(q_0, q_1, q_2) = (1, 0, 0)$, we have the 4-velocity vector $v^\mu = (1, 0, 0, 0)$ of preferred observers that are all spatially at rest in spacetime for $\varepsilon = 1$, while for $\varepsilon =0$, we find $v^\mu = (1, 0, 0, -1)$ for the null geodesics that move opposite to the direction of wave propagation. We recall that with $(q_0, q_1, q_2) = (0, 0, 0)$, we find the null geodesics that indicate the direction of propagation of the wave. 

In the wave scenario, the solution for the 4-velocity vector  of the future-directed timelike and null geodesics is thus given by
\begin{equation}\label{L18}
v^\mu = \left(\frac{dt}{d\eta}, q_1\, \mathfrak{u}^{-2 s_1}, q_2 \,\mathfrak{u}^{-2 s_2}, \frac{dz}{d\eta}\right)\,,
\end{equation}
where $dt/d\eta$ and $dz/d\eta$ are given by Eqs.~\eqref{L16} and~\eqref{L17}, respectively. As measured by the preferred observers, the corresponding 4-velocity vector is
\begin{equation}\label{L19}
v^{\hat \alpha} = v_\mu e^{\mu}{}_{\hat \alpha}\,;
\end{equation}
that is, 
\begin{equation}\label{L20}
v^{\hat \alpha} = \left(\frac{dt}{d\eta}, q_1\, \mathfrak{u}^{- s_1}, q_2 \,\mathfrak{u}^{- s_2}, \frac{dz}{d\eta}\right)\,
\end{equation}
using the tetrad frame field given in Eq.~\eqref{L6}. We can write this result in the form
\begin{equation}\label{L21}
v^{\hat \alpha} = \Gamma ( 1, \hat{\bm{V}})\,,
\end{equation}
where $\Gamma := dt/d\eta$ and $\hat{\bm{V}}$ is the velocity of the free test particle or null ray as measured by the reference observers. For timelike geodesics, $\eta = \tau$ and $\Gamma = dt/d\tau$ is the Lorentz factor.

If $q_1 = q_2 = 0$, there is only motion along the $z$ direction; then, for timelike geodesics $\varepsilon = 1$ and  we have
 \begin{equation}\label{L22}
\Gamma = \frac{1+q_0^2}{2q_0}\,, \qquad  \hat{\bm{V}} = \left( 0, 0, \frac{1-q_0^2}{1+q_0^2}\right)\,, 
\end{equation}
so that there is motion in the direction of the wave for $0 < q_0 < 1$ and in the opposite direction for $q_0 > 1$,  while for null geodesics $\varepsilon = 0$ and Eq.~\eqref{L20} implies
 \begin{equation}\label{L23}
\Gamma = q_0/2\,, \qquad  \hat{\bm{V}}|_{\rm null} = ( 0, 0, -1)\,. 
\end{equation}
Henceforth, we assume that $q_1 \ne 0$ and $q_2\ne 0$. The remaining cases, namely,  ($q_1 = 0$, $q_2\ne 0$) and  ($q_1 \ne 0$, $q_2 = 0$), can be straightforwardly worked out along the lines indicated below.


\subsection{Timelike Geodesics}

Inspection of Eq.~\eqref{L5} reveals that for $3\pi/4 < \theta < 5 \pi/4$, we have $s_1 < 0$ and $s_2 > 0$, while for $5\pi/4 < \theta < 7 \pi/4$, we have $s_1 > 0$ and $s_2 < 0$. Otherwise, $s_1 > 0$ and $s_2 > 0$. We do not consider the situations where $s_1$ (or $s_2$) is equal to $0$ or $1$, since the spacetime is flat in these cases. 

 We are interested in the behavior of the future-directed timelike geodesics as $t \to \infty$. The fiducial observers are all fixed in space; hence, as $t\to \infty$, $\mathfrak{u} \to \infty$. Consider first the situation where 
 $(s_1 > 0, s_2 < 0)$ or $(s_1 < 0,  s_2 > 0)$; then, it follows from Eqs.~\eqref{L20} and~\eqref{L21} that as $\mathfrak{u} \to \infty$, $\Gamma \to \infty$ and 
 \begin{equation}\label{L24}
 \hat{\bm{V}} \to ( 0, 0, 1)\,. 
\end{equation}
It is remarkable that timelike geodesics all line up in the direction of wave propagation and the speed of the jet asymptotically approaches the speed of light. This result is physically meaningful, since the jet properties have been invariantly defined. 

Let us next consider the situation where $s_1 > 0$ and $s_2 > 0$. It turns out that as $\mathfrak{u} \to \infty$, we find  in this case  
 \begin{equation}\label{L25}
\Gamma \to \frac{1+q_0^2}{2q_0}\,, \qquad  \hat{\bm{V}} \to \left( 0, 0, \frac{1-q_0^2}{1+q_0^2}\right)\,. 
\end{equation}
This is a moderate form of the jet along the direction of the wave for $0 < q_0 < 1$ and in the opposite direction for $q_0 > 1$. 


\subsection{Null Geodesics}

In this case, $\varepsilon = 0$ and $\eta$ is an affine parameter along the world line. For $(s_1 > 0, s_2 < 0)$ or $(s_1 < 0,  s_2 > 0)$, we find that as $\mathfrak{u} \to \infty$, $\Gamma \to \infty$ and 
 \begin{equation}\label{L26}
 \hat{\bm{V}}|_{\rm null} \to ( 0, 0, 1)\,, 
\end{equation}
while for $s_1 > 0$ and $s_2 > 0$, $\Gamma \to q_0/2$ and 
 \begin{equation}\label{L27}
 \hat{\bm{V}}|_{\rm null} \to ( 0, 0, -1)\,. 
\end{equation}


\subsection{$\mathfrak{u} \to 0$}

It is interesting to determine what happens to timelike and null geodesics on approach to the singularity at $\mathfrak{u} = 0$. Inspection of Eqs.~\eqref{L20} and~\eqref{L21} reveals that for all relevant values of $s_1$ and $s_2$, as $\mathfrak{u} \to 0$, $\Gamma \to \infty$ and 
\begin{equation}\label{L28}
 \hat{\bm{V}}|_{\mathfrak{u} \to 0} \to ( 0, 0, 1)\,. 
\end{equation}

The general conclusion that we draw from our detailed consideration of cosmic jet formation in the plane wave spacetime of Eq.~\eqref{L1} is that in addition to the main jet that develops in the direction of wave propagation, there is in general a \emph{counterjet} moving in the opposite direction. The wave scenario for cosmic jet formation in plane wave spacetimes first came to light in~\cite{Bini:2014esa} and was elaborated and extended to nonplanar twisted gravitational waves in~\cite{Bini:2017qnd, Bini:2018gbq, Firouzjahi:2019qmy}. In the latter case, the nonuniformity of the wave front results in cosmic jets that are tilted with respect to the direction of wave propagation.  Further interesting work in connection with the wave scenario is contained in ~\cite{Tucker:2016wvt, Tucker:2018xle}. More recently, the wave scenario has been verified in the general case of elliptically polarized plane gravitational wave spacetimes~\cite{Molaei:2025udr}. For a deeper understanding of the wave scenario, we work out the solution of the scalar wave equation in a more general setting involving elliptically polarized gravitational plane waves in the next section. 


\section{Wave Scenario: Scalar Field}

We consider a generalization of Eq.~\eqref{L1} that represents elliptically polarized gravitational plane waves propagating in the $z$ direction, namely, 
\begin{equation}\label{E1}
ds^2 = - dt^2 + dz^2 + P^2(\mathfrak{u}) dx^2 + 2 S(\mathfrak{u}) dx dy +Q^2(\mathfrak{u}) dy^2\,,
\end{equation}
where $P$, $Q$ and $S$ are functions of the retarded null coordinate $\mathfrak{u} = t-z$. The system of coordinates is admissible if $P(\mathfrak{u}) \ne 0$ and
\begin{equation}\label{E2}
\Delta(\mathfrak{u}) := P^2(\mathfrak{u}) \,Q^2(\mathfrak{u}) - S^2(\mathfrak{u}) >0\,.
\end{equation}
For the plane wave of the previous section, $P = \mathfrak{u}^{s_1}$, $Q = \mathfrak{u}^{s_2}$ and $S = 0$. The Ricci-flat condition in the case of Eq.~\eqref{E1} results in an ordinary differential equation involving $P$, $Q$ and $S$~\cite{Molaei:2025udr}. We are interested in the general solution of the scalar wave equation~\eqref{S3} in this spacetime, where $\det(g_{\mu \nu}) = - \Delta$ and the inverse metric is given by
\begin{equation}
\label{E3}
(g^{\mu \nu})= \frac{1}{\Delta}\,\left[
\begin{array}{cccc}
- \Delta & 0 & 0 & 0\cr
0 & Q^2 & -S & 0\cr
0 & - S & P^2 & 0\cr
0 & 0 & 0 & \Delta \cr
\end{array}
\right]\,.
\end{equation}

The scalar wave equation~\eqref{S3} for the gravitational field under consideration can be expressed as
\begin{equation}\label{E4}
 - \Delta^{\frac{1}{2}}\,\frac{\partial}{\partial t}\left(\Delta^{\frac{1}{2}}\,\frac{\partial \Phi}{\partial t}\right) + \Delta^{\frac{1}{2}}\,\frac{\partial}{\partial z}\left(\Delta^{\frac{1}{2}}\,\frac{\partial \Phi}{\partial z}\right) +  Q^2\, \frac{\partial^2 \Phi}{\partial x^2} -2 S\,\frac{\partial^2 \Phi}{\partial x \partial y}  + P^2\, \frac{\partial^2 \Phi}{\partial y^2} - \frac{\Delta}{\ell^2} \Phi = 0\,.
\end{equation}
Based on the existence of the Killing vectors $\partial_x$, $\partial_y$ and  $\partial_t + \partial_z$, we assume a solution of the form 
\begin{equation}\label{E5}
 \Phi = e^{i  C_1 x + i  C_2 y + i  C_3 (t+z)}\, \Psi(\mathfrak{u})\,,
\end{equation}
where the real part of this expression is expected to be physically significant. Substitution of this ansatz in Eq.~\eqref{E4} leads to a great deal of simplification and we end up with a first-order ordinary differential equation. That is, Eqs.~\eqref{E4} and~\eqref{E5} imply
\begin{equation}\label{E6}
C_3 \left(4\,\frac{d\Psi}{d\mathfrak{u}} + \frac{1}{\Delta}\frac{d\Delta}{d\mathfrak{u}}\,\Psi\right) = i\, (\ell^{-2} + \mathfrak{D})\,\Psi\,,
\end{equation}
where 
\begin{equation}\label{E7}
 \mathfrak{D}(\mathfrak{u}) := \frac{C_1^2 Q^2 - 2C_1C_2S + C_2^2P^2}{\Delta}\,.
\end{equation}
Clearly, there is a solution provided $C_3 \ne 0$. With this assumption, we find 
\begin{equation}\label{E8}
4\, \frac{1}{\Psi}\,\frac{d\Psi(\mathfrak{u})}{d\mathfrak{u}} = -\frac{1}{\Delta}\,\frac{d\Delta}{d\mathfrak{u}} + \frac{i}{C_3} (\ell^{-2} + \mathfrak{D})\,.
\end{equation}
Therefore, 
\begin{equation}\label{E9}
\Psi = \Delta^{-\frac{1}{4}}\,\exp{\left(\frac{i\, \mathfrak{u}}{4\,C_3\,\ell^2}\right)}\, \exp{\left(\frac{i}{4\,C_3}\, \int^{\mathfrak{u}} \mathfrak{D}(\mathfrak{u'})\,d\mathfrak{u'}\right)}\,
\end{equation}
and the complete solution is then a general linear combination of 
\begin{equation}\label{E10}
 \Phi|_{(C_1, C_2, C_3)} = \Delta^{-\frac{1}{4}}\,e^{i  C_1 x + i  C_2 y}\,\exp{\left(i \, C_3 \,(t+z) +\frac{i\, \mathfrak{u}}{4\,C_3\,\ell^2}\right)}\, \exp{\left(\frac{i}{4\,C_3}\, \int^{\mathfrak{u}} \mathfrak{D}(\mathfrak{u'})\,d\mathfrak{u'}\right)}\,
\end{equation}
for different values of the constants $C_1$, $C_2$ and $C_3 \ne 0$. 

Consider, for instance, the case where $C_1 = C_2 = 0$; then, $ \mathfrak{D} = 0$  and the massless ($m = 0,~ \ell = \infty$) scalar field $\Phi$ in the form $\Delta^{\frac{1}{4}}\,\Phi$ only involves the advanced null coordinate $t + z$, which indicates the presence of the counterjet propagating at the speed of light in the opposite direction of propagation of the plane wave, in agreement with Eq.~\eqref{L23}. For $m \ne 0$, the massive scalar field in this case would similarly involve both the retarded and the advanced null coordinates, namely, $C_3 \,(t+z) +\mathfrak{u}/(4\,C_3\,\ell^2)$, and we expect that the complete solution in the WKB limit would be in general agreement with Eq.~\eqref{L22}.

Consider next the general case where $C_1 \ne 0$ and $C_2 \ne 0$. Let us note that for the plane wave solution of the previous section $\Delta = \mathfrak{u}^{2(s_1 + s_2)}$ and $\mathfrak{D} = C_1^2 \mathfrak{u}^{-2s_1} + C_2^2 \mathfrak{u}^{-2s_2}$. For $\mathfrak{u} \to \infty$ and $s_1 > 0$ and $s_2 > 0$, we find $\mathfrak{D} \to 0$; hence, we asymptotically approach the situation already encountered in Eqs.~\eqref{L22} and~\eqref{L23}, in agreement with Eqs.~\eqref{L25} and~\eqref{L27}. On the other hand, for  $(s_1 > 0, s_2 < 0)$ or $(s_1 < 0,  s_2 > 0)$, $\mathfrak{D} \to \infty$ as $\mathfrak{u} \to \infty$ and the result is a cosmic jet in the direction of wave propagation irrespective of the mass of the scalar field, in agreement with Eqs.~\eqref{L24} and~\eqref{L26}.  

The behavior of the massive or massless scalar field on approach to the singularity $\mathfrak{u} \to 0$ is dominated by the fact that $\mathfrak{D} \to \infty$ for all three cases, namely,   $(s_1 > 0, s_2 > 0)$, $(s_1 > 0, s_2 < 0)$ and $(s_1 < 0,  s_2 > 0)$; therefore, a cosmic jet develops in the direction of wave propagation as in Eq.~\eqref{L28}. 

These significant scalar wave  results help to strengthen the physical basis of the wave scenario.


\section{DISCUSSION}

We have briefly described the collapse and wave scenarios for the formation of cosmic jets, where, relative to fiducial observers that are all spatially at rest  in time-dependent gravitational fields in GR,  timelike geodesics asymptotically line up and their speeds can approach the speed of light.   Timelike (null) geodesics can be obtained as the WKB limit of massive (massless) neutral scalar wave equation. For further insight into the jet formation process, we have investigated the propagation of scalar waves in spacetimes in which the collapse and wave scenarios can theoretically take place.   

Cosmic jets in GR only involve the gravitational interaction. Astrophysical jets, on the other hand,  involve persistent bipolar plasma outflows that require an appropriate magnetohydrodynamic (MHD) environment. Is there a connection between cosmic jets in GR and astrophysical jets? The answer is not known at the present time.


\appendix

\end{document}